\documentclass[aps,prx,reprint,runinaddress,superscriptaddress,amsmath,amssymb,floatfix]{revtex4-1}
\usepackage{wasysym}
\usepackage{times}
\usepackage[varg]{txfonts}
\usepackage{mathrsfs}
\usepackage{bm}

\usepackage{color}
\usepackage{graphicx}
\usepackage{hyperref}
\usepackage{booktabs}
\usepackage{natbib}
\usepackage{braket}
\usepackage{overpic}

\renewcommand{\vec}[1]{\boldsymbol{#1}}
\newcommand{\vhat}[1]{\vec{\hat{#1}}}
\newcommand{\mat}[1]{\vec{#1}}

\newcommand{\tss}[1]{\textsubscript{#1}}
\newcommand{\tsp}[1]{\textsuperscript{#1}}

\newcommand{\byzo}{Ba\tss{3}Yb\tss{2}Zn\tss{5}O\tss{11}}

\newcommand{\yb}{Yb\tsp{3+}}
\newcommand{\K}{\ {\rm K}}
\newcommand{\mK}{\ {\rm mK}}

\newcommand{\trp}[1]{{#1}^{\intercal}}

\usepackage[nolist,nohyperlinks]{acronym}  
\newacro{BYZO}[BYZO]{\byzo{}}
\newacro{DM}[DM]{{Dzyaloshinskii-Moriya}}
\newacro{BP}[BP]{breathing pyrochlore}
\newacro{INS}[INS]{inelastic neutron scattering}
\newacro{AIAO}[AIAO]{all-in/all-out}

\newcommand{\meV}{\ {\rm meV}}
\newcommand{\T}{\ {\rm T}}
\newcommand{\hc}{{\rm h.c.}}

\definecolor{cblue}{RGB}{33,145,140}
\hypersetup{colorlinks=true,linkcolor=cblue,citecolor=cblue,urlcolor=cblue}

\begin{document}

\title{Behavior of the breathing pyrochlore lattice \byzo{} in applied magnetic field}

\author{J. G. Rau}
\affiliation{Department of Physics and Astronomy, University of Waterloo, Ontario, N2L 3G1, Canada}
\affiliation{Max-Planck-Institut f\"ur Physik komplexer Systeme, 01187 Dresden, Germany}

\author{L. S. Wu}
\affiliation{Neutron Scattering Division, Oak Ridge National Laboratory, Oak Ridge, TN-37831, USA}

\author{A. F. May}
\affiliation{Materials Science \& Technology Division, Oak Ridge National Laboratory, Oak Ridge, TN-37831, USA}

\author{A. E. Taylor}
\affiliation{Neutron Scattering Division, Oak Ridge National Laboratory, Oak Ridge, TN-37831, USA} 

\author{I-Lin Liu}
\affiliation{Center for Nanophysics and Advanced Materials, Department of Physics University of Maryland
}
\affiliation{NIST Center for Neutron Research, National Institute of Standards and Technology}

\affiliation{Department of Materials Science and Engineering, University of Maryland, College Park, MD 20742, USA.}

\author{J. Higgins}
\affiliation{Center for Nanophysics and Advanced Materials, Department of Physics University of Maryland
}

\author{N. P. Butch}
\affiliation{Center for Nanophysics and Advanced Materials, Department of Physics University of Maryland
}
\affiliation{NIST Center for Neutron Research, National Institute of Standards and Technology}

\author{K. A. Ross}
\affiliation{Department of Physics, Colorado State University, 200 W. Lake St., Fort Collins, CO 80523-1875, USA}
\affiliation{Quantum Materials Program, Canadian Institute for Advanced Research, MaRS Centre, West Tower 661 University Ave., Suite 505, Toronto, ON, M5G 1M1, Canada}

\author{H. S. Nair}
\affiliation{Department of Physics, Colorado State University, 200 W. Lake St., Fort Collins, CO 80523-1875, USA}

\author{M. D. Lumsden}
\affiliation{Neutron Scattering Division, Oak Ridge National Laboratory, Oak Ridge, TN-37831, USA}

\author{M. J. P. Gingras}
\affiliation{Department of Physics and Astronomy, University of Waterloo, Ontario, N2L 3G1, Canada}
\affiliation{Quantum Materials Program, Canadian Institute for Advanced Research, MaRS Centre, West Tower 661 University Ave., Suite 505, Toronto, ON, M5G 1M1, Canada}
\affiliation{Perimeter Institute for Theoretical Physics, Waterloo, Ontario, N2L 2Y5, Canada} 

\author{A. D. Christianson}
\affiliation{Neutron Scattering Division, Oak Ridge National Laboratory, Oak Ridge, TN-37831, USA}

\affiliation{Materials Science \& Technology Division, Oak Ridge National Laboratory, Oak Ridge, TN-37831, USA}

\date{\today}

\begin{abstract}
The breathing pyrochlore lattice material \byzo{} exists in the nearly decoupled limit, in contrast to most other well-studied breathing pyrochlore compounds. As a result, it constitutes  a useful platform to benchmark theoretical calculations of exchange interactions in insulating  Yb$^{3+}$ magnets. Here we study \byzo{} at low temperatures in applied magnetic fields as a further probe of the physics of this model system.  Experimentally, we consider the behavior of polycrystalline samples of \byzo{} with a combination of inelastic neutron scattering and heat capacity measurements down to 75 mK and up to fields of 10 T.  Consistent with previous work, inelastic neutron scattering finds a level crossing near 3 T, but no significant dispersion of the spin excitations is detected up to the highest applied fields.  Refinement of the theoretical model previously determined at zero field can reproduce much of the inelastic neutron scattering spectra and specific heat data.  A notable exception is a low temperature peak in the specific heat at $\sim 0.1 \K$.  This may indicate the scale of interactions between tetrahedra or may reflect undetected disorder in \byzo{}. \footnote{This manuscript has been authored by UT-Battelle, LLC under Contract No. DE-AC05-00OR22725 with the U.S. Department of Energy.  The United States Government retains and the publisher, by accepting the article for publication, acknowledges that the United States Government retains a non-exclusive, paid-up, irrevocable, world-wide license to publish or reproduce the published form of this manuscript, or allow others to do so, for United States Government purposes.  The Department of Energy will provide public access to these results of federally sponsored research in accordance with the DOE Public Access Plan (http://energy.gov/downloads/doe-public-access-plan).}
\end{abstract}

\maketitle

\section{Introduction}

Breathing pyrochlore lattice materials have recently emerged as an interesting route to study various aspects of frustrated magnetism~\cite{okamoto-2013-breathing,okamoto-2015-breathing,neel_2015,kimura-2014-breathing,Rau_2016,lee_2016,haku2016,Park2016,Okamoto_2018,savary_2016,Okamoto_2018,pokharel_2018}. Examples include the potentially enhanced stability of quantum spin ice state~\cite{savary_2016}, as well as the appearance of a ``Weyl magnon", a bosonic analog of a Weyl fermion~\cite{li_2016,jian_2018,ezawa_2018}, which hosts the associated magnon arc surface states~\cite{li_2016}. 

\begin{figure}[t]
    \centering
    \begin{overpic}[width=0.9\columnwidth]{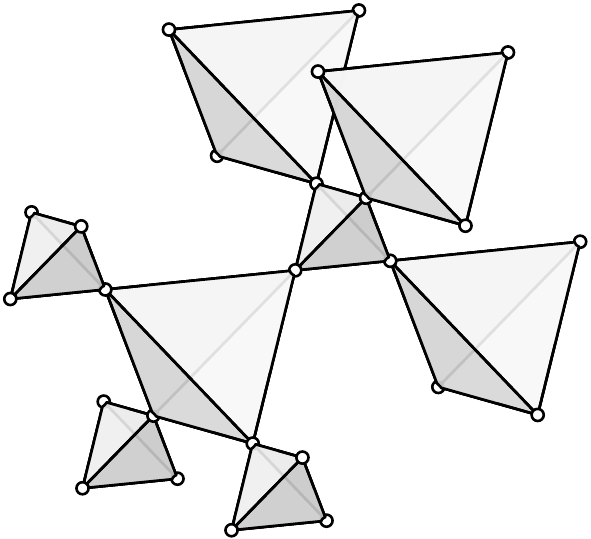}
    \put(57,40){\large $d$}
    \put(32,47){\large $d'$}
    \end{overpic}
\caption{\label{fig:bplattice}
Illustration of the breathing pyrochlore lattice, as formed by the {Yb}\textsuperscript{3+} ions on the vertices of large and small tetrahedra in \byzo{}.  The small and large tetrahedra have bonds of length $d$ and $d'$ (respectively), defining the breathing ratio $d'/d$.}
\end{figure}

The breathing pyrochlore lattice consists of a three dimensional network of corner sharing tetrahedra which alternate in size between large and small (see Fig.~\ref{fig:bplattice}). The different bond lengths provide a natural mechanism for a separation of energy scales between the short (length $d$) and long (length $d'$) bonds. By tuning the relative importance of these interactions, one can tune between decoupled tetrahedra and the full pyrochlore lattice.  Known examples of breathing pyrochlore materials have been found to lie at the extremes of these behaviors, realizing either near equal couplings~\cite{okamoto-2013-breathing,okamoto-2015-breathing,Okamoto_2018,pokharel_2018}, or having weakly coupled, nearly non-interacting tetrahedra~\cite{kimura-2014-breathing}.

The case of strongly coupled tetrahedra appears in several Cr-based spinels such as Li(In,Ga)Cr$_4$O$_8$~\cite{okamoto-2013-breathing,okamoto-2015-breathing} and Li(In,Ga)Cr$_4$S$_8$ \cite{Okamoto_2018,pokharel_2018}. In these materials the difference between the size of the large and small tetrahedra is small, with ``breathing ratios'' in the range $ 1.05 \lesssim d'/d  \lesssim 1.1$.  The strength of the interactions between the (nominally) small tetrahedra yields behavior closer to the full pyrochlore lattice, instead of the decoupled limit, resulting in a transition to a magnetic ground state in most of these materials~\cite{okamoto-2013-breathing,okamoto-2015-breathing,Okamoto_2018,pokharel_2018}.

In contrast, the compound \byzo{} realizes the opposite limit of the breathing pyrochlore lattice, with a large breathing ratio of $d'/d \sim 2$~\cite{kimura-2014-breathing}. The lack of evidence of magnetic ordering in this compound, combined with a residual entropy of $\sim \log{2}$ per small tetrahedron strongly suggests that the inter-tetrahedron coupling is weak~\cite{kimura-2014-breathing}, with the material being described by essentially decoupled tetrahedral units~\cite{Rau_2016,Park2016,haku2016}.
The physics of these decoupled tetrahedra has been firmly established by several inelastic neutron scattering (INS) studies~\cite{Rau_2016,haku2016,Park2016}. These investigations, and the associated theoretical models, have revealed that the intra-tetrahedron physics originates from dominant anti-ferromagnetic Heisenberg exchange~\cite{kimura-2014-breathing} coupled with an unusually strong \ac{DM} interactions~\cite{Rau_2016,haku2016,Park2016}. Such interactions yield a nearly non-magnetic doublet as ground state of each tetrahedron,
consistent with expectations from the residual entropy and the magnetic susceptibility~\cite{kimura-2014-breathing}. As a result of its experimental and theoretical tractability, \byzo{} serves an ideal model system for studying the microscopic mechanisms of anisotropic exchange interactions in rare-earth compounds~\cite{rau2018frustration}. The ability to unambiguously determine the exchange parameters in this compound~\cite{Rau_2016,haku2016,Park2016} has revealed several new features of rare-earth exchange, namely the emergent ``weak" anisotropy that can appear in ytterbium magnets composed of edge-sharing octahedra~\cite{rau2018frustration}.

Two key issues that remain unresolved in \byzo{} center around (1) the strength and importance of the inter-tetrahedron interactions and (2) the response of \byzo{} to applied magnetic fields. The later question is particularly interesting since the excited magnetic states should be significantly more susceptible to field tuning, in contrast to the nearly nonmagnetic ground state.  Some progress on these issues was reported in Refs.~[\onlinecite{haku2016},\onlinecite{Park2016}] which studied inelastic neutron scattering and magnetization as a function of applied field.  Further, evidence of physics beyond isolated tetrahedra may have been observed in zero field specific heat measurements, which reveal a broad peak at $\sim 63 \mK$~\cite{haku2016}. Whether this release of entropy is related to inter-tetrahedron interactions~\cite{Rau_2016,haku2016}, or is the result of extrinsic effects, such as structural disorder~\cite{haku2016},
has not yet been established.

The goal of this paper is to explore these questions in more detail using both INS and heat capacity measurements in applied magnetic fields.  This combination of measurements is particularly powerful as the INS measurements are sensitive to the overall excitation spectrum, including wave vector dependence, while the specific heat is much more sensitive to very low energy features. In particular, to explore the low-temperature feature at $\sim 63 \mK$~ in detail, we go beyond previous studies and track its dependence on magnetic field~\cite{haku2016}. We support our interpretation of these experimental results with theoretical modeling, building on the results described in Ref. [\onlinecite{Rau_2016}].  Our experimental INS results are generally consistent with those of Ref. [\onlinecite{Park2016}], but we find that, to obtain good agreement with model calculations, careful consideration of the direction of the applied field in relation to the experimental geometry is essential. The resulting single tetrahedron model provides a good description of the INS data as well as the heat capacity at low fields. Two notable exceptions appear: first are the low-temperature features present in the heat capacity for $T \lesssim 1 \K$, which are not reproduced by the model and second, are some aspects of the specific heat and INS data at high fields. We speculate that the low-temperature features either provide an indication of the scale of interactions between the small tetrahedra or represents and extrinsic effect such as disorder, as in the zero-field case.  The high-field case is less clear, but we discuss some possible explanations of the aforementioned deviations.

\begin{figure*}[htp]
  \centering
  \includegraphics[width=0.98\textwidth]{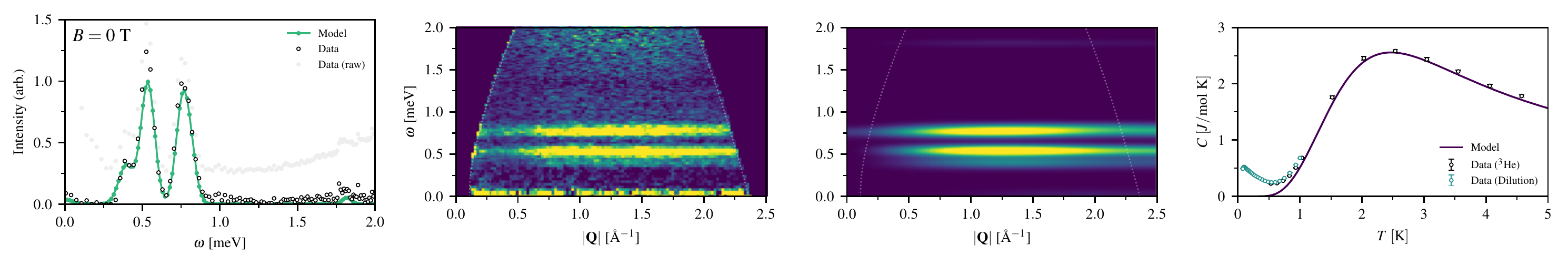}
  \includegraphics[width=0.98\textwidth]{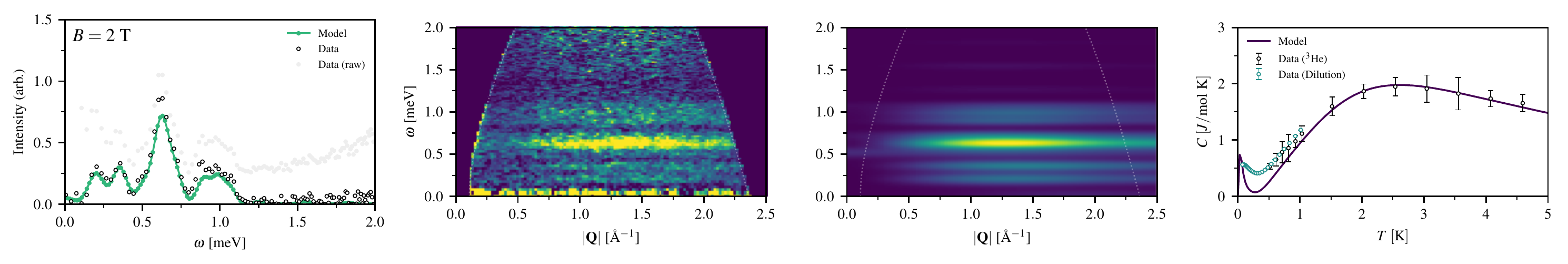}
  \includegraphics[width=0.98\textwidth]{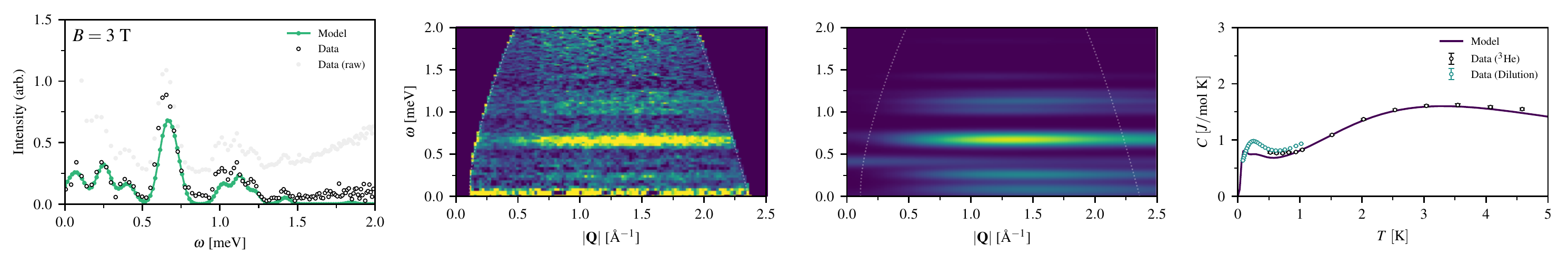}
  \includegraphics[width=0.98\textwidth]{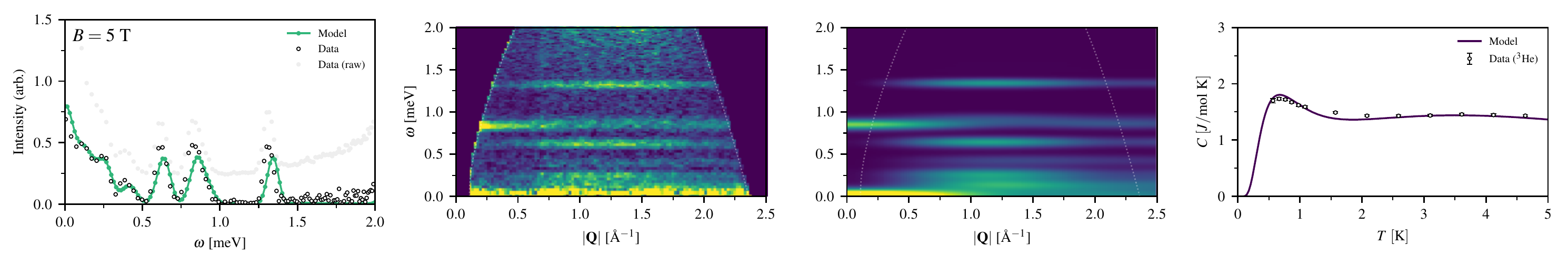}
  \includegraphics[width=0.98\textwidth]{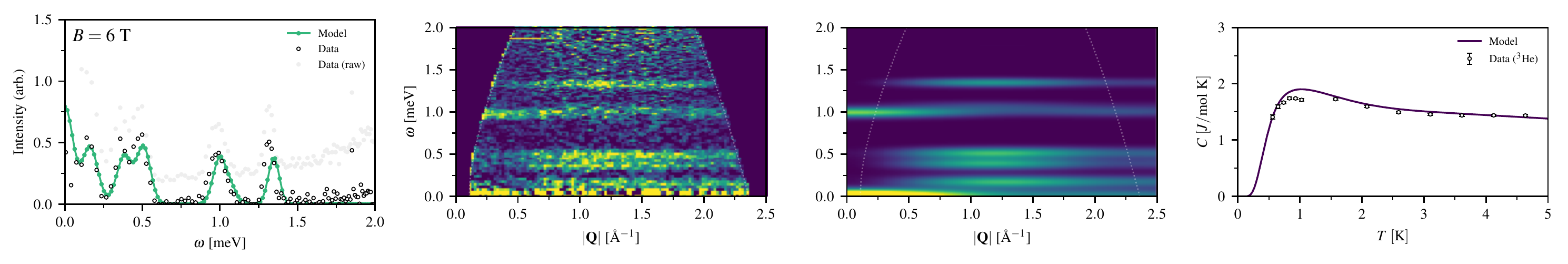}
  \includegraphics[width=0.98\textwidth]{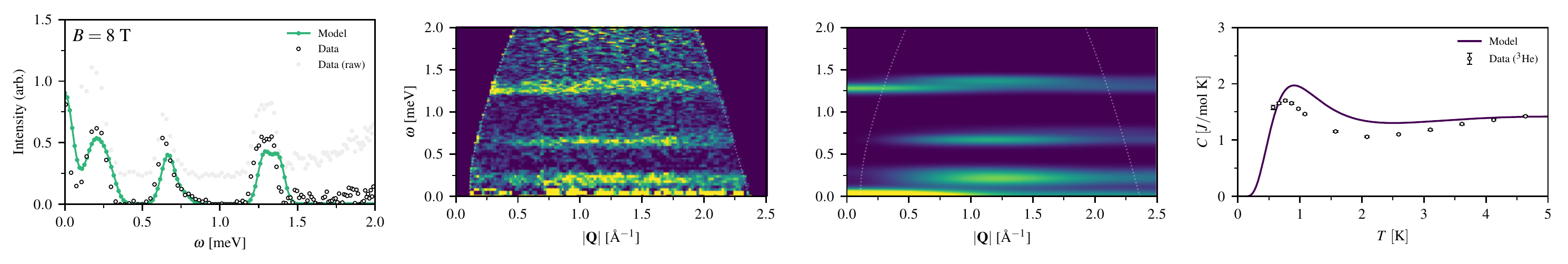}    
  \includegraphics[width=0.98\textwidth]{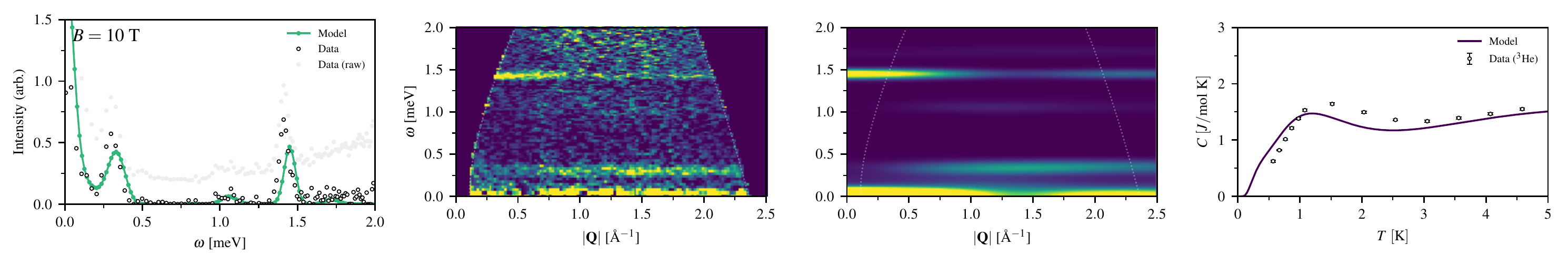}
  \caption{\label{fig:results}
Experimental and theoretical results in magnetic field from $0\T$ to $10\T$. Each row of panels shows results for a single magnetic field strength.  The neutron scattering data displayed in the figure were collected at $T = 0.1 \K$.  The first panel shows a cut of the INS data integrated over all available data in the range $0 ~{\mbox{\AA}}^{-1} < |\vec{Q}| < 2.5 ~{\mbox{{\mbox{\AA}}}}^{-1}$ (symbols), with and without background, as well as the associated model calculation (line).  The second and third columns display the full INS spectrum, $I(Q,\omega)$ and the associated refined model calculation.  The fourth panel shows the experimental (symbols) and model calculation (line) of the temperature dependent specific heat.}
\end{figure*}

\section{Experimental results}
\subsection{Specific heat}
\begin{figure}[tp]
    \centering
    \includegraphics[width=0.9\columnwidth]{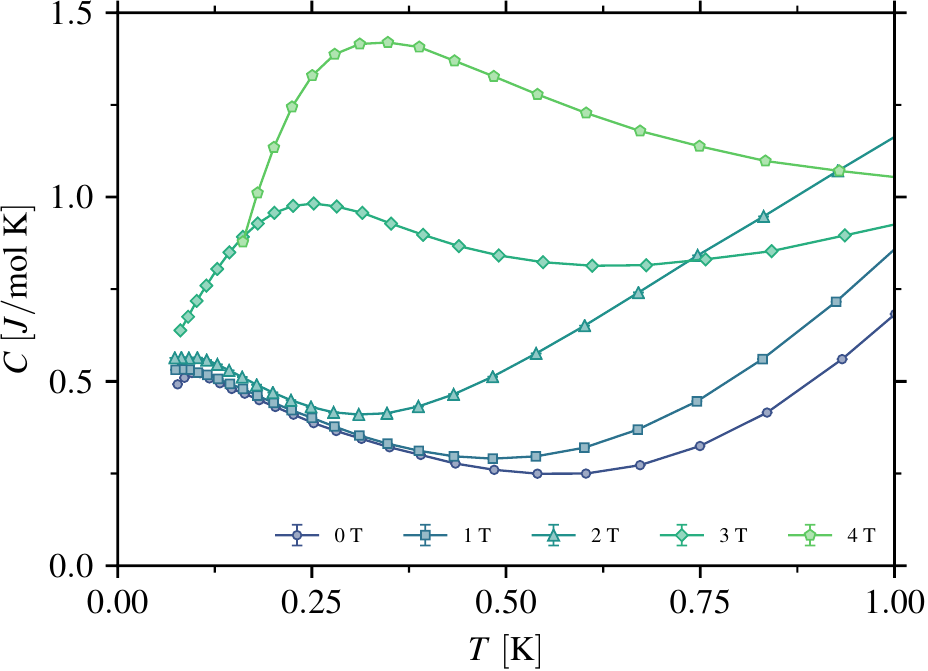}
\caption{\label{fig:specific-heat-low}
Field dependence of the low temperature ($T < 1\K$) specific heat (solid lines are guides to the eye). In this temperature range, the single-tetrahedron model described in the text no longer provides a good description of the data.
}
\end{figure}
Specific heat measurements were performed on pelletized polycrystalline samples of \byzo{} prepared as described in Ref.~[\onlinecite{Rau_2016}] using a $^3$He insert and a dilution refrigerator insert, in a Quantum Design Physical Property Measurement System (PPMS). The specific heat as function of temperature ($0.5 \K \leq T \leq 5 \K$)  for several applied magnetic fields is displayed in the rightmost column of Fig. \ref{fig:results} and in Fig. \ref{fig:specific-heat-low}.  The phonon contribution the specific heat has been removed by subtraction of the iso-structural, non-magnetic compound Ba$_3$Lu$_2$Zn$_5$O$_{11}$ for the $^3$He insert measurements, but not for the measurements in the dilution refrigerator due to the negligible phonon contribution at low temperatures.
A prominent feature of the data is the evolution with field of the specific heat at low temperatures.  As will be discussed in more detail below, this is consistent with a lowering of the energy of excited states as a function of applied field. 

To gain further insight into the low energy spin dynamics as a function of magnetic field, the specific heat was measured down to $\sim 75 \mK$ for several field strengths (Fig. \ref{fig:specific-heat-low}).  At zero field, a peak centered near $\sim 100 \mK$ is observed.  Although the measurements presented here have a minimum temperature of 75 mK, this feature likely corresponds to the peak at $\sim 63 \mK$ observed by Ref.~[\onlinecite{haku2016}] as no additional upturn is observed in the data presented here as would be expected if the feature at $\sim 63 \mK$ was to be manifest in our measurements down to this temperature. The observed difference in the position of this peak may indicate that an extrinsic effect such as structural disorder 
 may play some role
 (i.e. a slight difference between the sample ``quality'' of Ref.~[\onlinecite{haku2016}] and our own).
 However, no evidence of structural disorder was reported in previous neutron diffraction studies leaving the influence of disorder, if present, unclear~\cite{Rau_2016,Park2016}.  Note that due the nearly non-magnetic ground doublets of each tetrahedron, hyperfine contributions are suppressed and are only expected to be significant for $T\lesssim 50\mK$. The low temperature specific heat changes smoothly with field up to $2\T$.  At higher fields, more significant changes are observed.  As described in Ref.~[\onlinecite{Park2016}] and in Sec. \ref{theory} below, these changes are associated with level crossings of the single tetrahedron ground state with one of the excited states.
 
\subsection{Neutron scattering}
Neutron scattering measurements were performed at the Disk Chopper Spectrometer (DCS) at NCNR, NIST, with incident neutron energy $E_i=3.27 \rm meV$ ($\lambda_i=5.0 \rm {\mbox{\AA}}$) \cite{DCS}. A dilution refrigerator insert that can access temperatures as low as 70 mK was used with a 10 T magnet. The polycrystalline samples of \byzo{} used here were synthesized by solid-state reaction and characterized as described in Ref.~\cite{Rau_2016}. 

To remove the background, we leverage the discrete nature of the sharp, essentially dispersionless single tetrahedron energy levels and their strong dependence on magnetic field. For each $|\vec{Q}|$ and $\omega$, we assume that there is a field value (covering ten or so field strengths in the range  $0 \T \leq B \leq 10\T$) such that there is no magnetic contribution to the intensity at this point and use this as the background to be subtracted from the raw data. More explicitly, for each $|\vec{Q}| \equiv Q$ and $\omega$, we take the background to be $I_{\rm 0}(Q,\omega) \equiv {\rm min}_{B} I(Q,\omega;B)$, where $I(Q,\omega;B)$ is the experimental intensity at field strength $B$. Note that this subtraction procedure ensures that the intensity always remains positive.

\begin{figure}[tp]
    \centering
    \includegraphics[width=0.9\columnwidth]{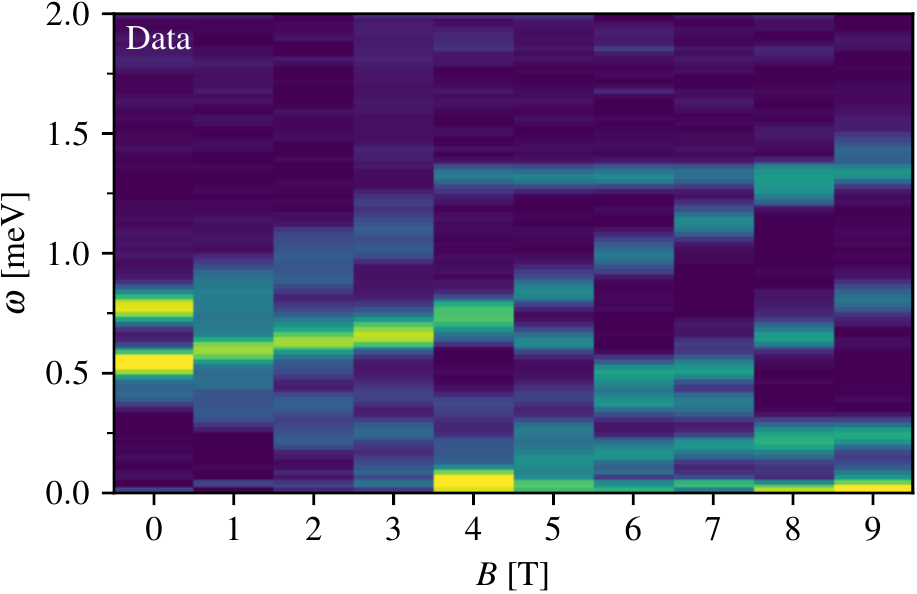}
\caption{\label{fig:spectrum}
Summary of the field dependence of the inelastic neutron scattering spectrum measured at $0.1\K$.  Given the flat, dispersionless modes we have integrated over the range of wave-vectors $0 ~{\mbox{\AA}}^{-1} \leq |\vec{Q}|\leq 2 ~{\mbox{\AA}}^{-1}$ (see Fig. \ref{fig:results} for the full spectrum). As discussed in the text, level crossings are visible near $B \sim 3-4\T$ and near $B \sim 8-9\T$.}
\end{figure}

A summary of the field dependence of the excitation spectrum measured at $0.1\K$ can be obtained by integrating over the range  $0~{\mbox{\AA}}^{-1} \leq |\vec{Q}|\leq 2 ~{\mbox{\AA}}^{-1}$, as shown in Fig. \ref{fig:spectrum}.  This data is consistent with previous studies of \byzo{}\cite{haku2016,Park2016,Rau_2016}. Further details are provided in Fig. \ref{fig:results} which shows the $|\vec{Q}|$ and $\omega$ dependence of the spectra at several applied fields. The spectrum at zero field comprises several non-dispersive modes. While the observed modes remain remain largely dispersionless up to $10\T$, the highest field probed in this study, these levels are reorganized dramatically as the field is raised. 

For small applied fields, one finds the spectral weight associated with the levels near $\sim 0.5 \meV$ is pushed to lower energy, a result of the splitting of the excited state levels, as the ground state doublet is nearly non-magnetic and only responds weakly to the field.  Near $B^{(1)}_c \sim 3-4 \T$, a more dramatic change occurs in the spectra reflecting a level crossing where one of the excited states becomes the ground state (see Sec. \ref{theory}). Note that due to the anisotropic nature of the spin Hamiltonian for this system~\cite{Rau_2016}, the precise location of a level crossing will depend on the direction of the field relative to the crystal axes. As these measurements are performed on powder samples, this then implies that there will be a distribution of such crossings and thus endowing some (``powder averaging'')  width to the observed transition.

Above this crossing ($B \gtrsim B^{(1)}_c$), the spectrum is qualitatively different, with additional modes that exhibit a significantly different $|\vec{Q}|$-dependence.  The majority of these modes continue to have a maximum intensity near $1.3 {\mbox{\AA}}^{-1}$.  However, for $B \gtrsim 5 \T$, a mode appears (starting at $\omega \sim 0.75 \meV$) which has a maximum intensity at low $|\vec{Q}|$ rather than at $1.3 {\mbox{\AA}}^{-1}$ and thus appears to be more consistent with an isolated spin flip rather than a collective mode of a tetrahedron.  As seen in Fig. \ref{fig:results}, this mode moves up in energy with increasing field. A second level crossing occurs at $B^{(2)}_c\sim 8-9\T$, near the high end of the field range considered here. As in the case of the first crossing, significant changes in the excited levels are observed.  However, given its proximity to our maximum field, we have not studied this crossing in the same detail.

\section{Theoretical results}\label{theory}
\subsection{Model}

We now briefly review the appropriate low-energy effective model and compare these results to the predictions for the parameters established in Refs.~[\onlinecite{Rau_2016,haku2016,Park2016}]. Given the large distance ratio, $d' / d \sim 2$ between the small ${\rm Yb}_4$ tetrahedra, we
follow the prior works and neglect any inter-tetrahedron interactions. Each \yb{} ion in this tetrahedron hosts an effective spin-1/2 degree of freedom, formed from the ${}^2 F_{7/2}$ ($J=7/2$) Hund's rule ground state manifold by the $C_{3v}$ crystal electric field environment. The other crystal field doublets are well-separated from the ground doublet relative to the scale of intra-tetrahedron interactions; the lowest lying state is at $\sim 38\meV$~\cite{haku-2015-breathing}. For each single-ion crystal field doublet we define a pseudo-spin operator
$\vec{S}_i$ which is related to the magnetic moment $\vec{\mu}_i$  through the $g$-factors, $g_z$ and $g_{\pm}$
\begin{equation}
\label{eq:moment}
    \vec{\mu}_i \equiv  \mu_B \left[ g_{\pm} \left(\vhat{x}_i S^x_i +  \vhat{y}_i S^y_i\right) + g_z \vhat{z}_i S^z_i\right],
\end{equation}
where $(\vhat{x}_i, \vhat{y}_i, \vhat{z}_i)$ are the local axes of tetrahedron site $i$~\cite{Rau_2016}. The symmetry of the ${\rm Yb}_4$ tetrahedra constrains the interactions between these pseudo-spins to be of the form
\begin{align}
\label{eq:model}
    & H_{\rm eff} \equiv \sum_{i=1}^4 \sum_{j<i} \left[
        J_{zz} S^z_i S^z_j - J_{\pm}\left(S^+_i S^-_j+S^-_i S^+_j\right)+\right.  \nonumber \\
        & J_{\pm\pm} \left(\gamma_{ij} S^+_i S^+_j+\hc \right)+ 
        \left. J_{z\pm}  \left(
            \zeta_{ij} \left[ S^z_i S^+_j+ S^+_i S^z_j \right]+ \hc
            \right) 
    \right],
\end{align}
where the bond dependent phases $\gamma_{ij}$ and $\zeta_{ij}$  are defined in Ref.~[\onlinecite{Rau_2016}]. Some aspects
of this model are better understood in a global basis for the pseudo-spins. If the global pseudo-spin operators
$\tilde{\vec{S}}_i$ are defined as 
\begin{equation}
  \tilde{\vec{S}}_i \equiv \vhat{x}_i S^x_i + \vhat{y}_i S^y_i+ \vhat{z}_i S^z_i,
\end{equation}
then the model of Eq.~(\ref{eq:model}) takes the form
\begin{equation}
\label{eq:model-two}
    H_{\rm eff} = \sum_{i=1}^4 \sum_{j<i} \trp{\tilde{\vec{S}}}_i \tilde{\mat{J}}_{ij} \tilde{\vec{S}}_j,
\end{equation}
where the global exchange matrices $\tilde{\mat{J}}_{ij}$ are defined as
\begin{align}
    \tilde{\mat{J}}_{12} &= \left(
        \begin{array}{ccc}
             J+K &+\frac{D}{\sqrt{2}} &+\frac{D}{\sqrt{2}}  \\
             -\frac{D}{\sqrt{2}} & J & \Gamma \\
             -\frac{D}{\sqrt{2}} & \Gamma & J
        \end{array}
    \right), & 
    \tilde{\mat{J}}_{13} &= \left(
        \begin{array}{ccc}
             J & -\frac{D}{\sqrt{2}} & \Gamma  \\
            +\frac{D}{\sqrt{2}} & J+K &+\frac{D}{\sqrt{2}} \\
             \Gamma & -\frac{D}{\sqrt{2}} & J
        \end{array}
    \right), \nonumber \\    
    \tilde{\mat{J}}_{14} &= \left(
        \begin{array}{ccc}
             J & \Gamma & -\frac{D}{\sqrt{2}}  \\
             \Gamma & J & -\frac{D}{\sqrt{2}} \\
            +\frac{D}{\sqrt{2}} &+\frac{D}{\sqrt{2}} & J+K
        \end{array}
    \right), & 
    \tilde{\mat{J}}_{23} &= \left(
        \begin{array}{ccc}
             J & -\Gamma &+\frac{D}{\sqrt{2}}  \\
             -\Gamma & J & -\frac{D}{\sqrt{2}} \\
             -\frac{D}{\sqrt{2}} &+\frac{D}{\sqrt{2}} & J+K
        \end{array}
    \right), \nonumber \\    
    \tilde{\mat{J}}_{24} &= \left(
        \begin{array}{ccc}
             J &+\frac{D}{\sqrt{2}} & -\Gamma  \\
             -\frac{D}{\sqrt{2}} & J+K &+\frac{D}{\sqrt{2}} \\
             -\Gamma & -\frac{D}{\sqrt{2}} & J
        \end{array}
    \right), & 
    \tilde{\mat{J}}_{34} &= \left(
        \begin{array}{ccc}
             J+K & -\frac{D}{\sqrt{2}} &+\frac{D}{\sqrt{2}}  \\
            +\frac{D}{\sqrt{2}} & J & -\Gamma \\
             -\frac{D}{\sqrt{2}} & -\Gamma & J
        \end{array}
    \right).\nonumber
\end{align}
The local parametrization of Eq.~(\ref{eq:model}) and this global
parametrization are related as
\begin{align}
J &= \frac{1}{3} \left(+4 J_{\pm}+2 J_{\pm\pm}+2 \sqrt{2} J_{z\pm}-J_{zz}\right), \nonumber \\
K &= \frac{2}{3} \left(-4 J_{\pm}+ J_{\pm\pm}+ \sqrt{2} J_{z\pm}+ J_{zz}\right), \nonumber \\
\Gamma &= \frac{1}{3} \left(-2 J_{\pm}-4 J_{\pm\pm}+2 \sqrt{2} J_{z\pm}-J_{zz}\right), \nonumber \\
D &= \frac{\sqrt{2}}{3} \left(-2 J_{\pm}+2 J_{\pm\pm}-\sqrt{2} J_{z\pm}-J_{zz}\right).
\end{align}

The four exchange parameters present in this model [Eq.(\ref{eq:model})], in addition to the $g$-factors in the moment $\vec{\mu}_i$ [Eq.(\ref{eq:moment})] were fit to thermodynamic and INS data in Refs.~[\onlinecite{Rau_2016},\onlinecite{haku2016},\onlinecite{Park2016}]. For example, 
Ref.~[\onlinecite{Rau_2016}] found 
\begin{align}
\label{eq:best-global-old}
  J &= +0.587 \meV, &
  K &= -0.014 \meV, \nonumber \\
  \Gamma &= -0.011 \meV, &
  D &= -0.165 \meV.
\end{align}
with $g_z = 3.07$ and $g_{\pm} = 2.36$. The results of Refs.~[\onlinecite{Rau_2016},\onlinecite{haku2016},\onlinecite{Park2016}] are quantitatively similar, but have some differences in  details; most significantly in the values for the $g$-factors. However, in all analyses, one finds to a good approximation these fitted parameters describe a Heisenberg antiferromagnet ($J>0$) supplemented with large (indirect) DM interaction~\cite{canals-lacroix-2008-ising} ($D<0$) and negligible symmetric anisotropies ($K \sim \Gamma \sim 0$). The appearance of this exchange regime in \byzo{} has recently been rationalized on the basis of a theoretical calculation~\cite{rau2018frustration}.


The spectrum of this single-tetrahedron model can be understood from the limit of a pure Heisenberg anti-ferromagnet where the states can be classified by the total spin of the four spins. The ground states consist of a doublet of singlets ($S_{\rm tot} = 0$), with the excited states being three triplets ($S_{\rm tot} = 1$) and a single high-energy quintet ($S_{\rm tot}=2$)~\cite{tsunetsugu-2001-singlet}. Adding indirect \ac{DM} strongly mixes the three triplets but preserves the doublet ground state and has no effect on the high lying quintet. The projection of a pseudo-spin $\tilde{\vec{S}}_i$ into the single \emph{tetrahedron} ground doublet states, denoted $\ket{\pm}$, takes the form $\bra{{\pm}}\tilde{\vec{S}}_i\ket{{\pm}} = \pm \lambda \vhat{z}_i$ with $\lambda \sim 0.1$ and $\bra{{\pm}} \tilde{\vec{S}}_i\ket{{\mp}}=0$. This lack of a dipole moment implies that the ground doublet will split only very weakly in small magnetic fields~\cite{Rau_2016}. Further, it puts strong constraints on the origin and scale of any inter-tetrahedron interactions; specifically at leading order these interactions are purely Ising in the doublet degrees of freedom and are suppressed by factors of $O(\lambda^2)$~\cite{Rau_2016}. These restrictions do \emph{not} apply to the higher lying states which, as we shall see, respond strongly to applied fields. 

\subsection{Refinement}
In a magnetic field, the single tetrahedron model takes the form
\begin{align}
  \label{eq:model-field}
  \sum_{i=1}^4 \left(\sum_{j<i} \trp{\tilde{\vec{S}}}_i \tilde{\mat{J}}_{ij} \tilde{\vec{S}}_j
     -  \mu_B \vec{B} \cdot \vec{\mu}_i\right),
\end{align}
where the moment $\vec{\mu}_i$ is defined in Eq.~(\ref{eq:model-two}). At zero field, the $g$-factors that describe in $\vec{\mu}_i$ projected in the single-ion crystal field doublet appear only in susceptibility and the intensities of the inelastic transitions. Neither of these probes are particularly suited to a precise determination of the $g$-factors. 

To refine the model for \byzo{}, we thus consider re-fitting the $g$-factors and exchange parameters, that we originally determined in Ref. [\onlinecite{Rau_2016}], using the INS spectrum in a magnetic field. One must account for the powder averaging carefully. We denote the intensity of a crystallite of the powder sample aligned to the laboratory frame as $I(\vec{Q},\omega;B\vhat{z})$, where $\vec{Q}$,$\omega$ are the wave-vector and energy and the applied field is chosen to be along the (laboratory) $\vhat{z}$ direction. For a grain of the powder sample that is rotated from the nominal axes by some rotation $\mat{R}$, the inelastic intensity at wave-vector $\vec{Q}$ in field $\vec{B}$ will then be given by $I(\trp{\mat{R}}\vec{Q},\omega;B\trp{\vec{R}}\vhat{z})$. Due to the detector geometry, only a narrow window of wave-vectors are collected, namely those with $\vhat{Q}$ approximately perpendicular to $\vhat{B}$~\footnote{The data analyzed here is for scattering angles $\pm3^{\circ}$ of the horizontal scattering plane and for $-30^\circ$ to $140^{\circ}$ within the horizontal scattering plane}.
Averaging over the grain orientations, $\mat{R}$, the powder-averaged intensity is then given 
\begin{equation}
    I(Q,\omega,B) \equiv \int d\vhat{B} \int_{\vhat{Q} \perp \vhat{B}} d\vhat{Q} \ I(Q\vhat{Q},\omega;B\vhat{B}).
\end{equation}
Note that due to the correlation between the field and wave-vector directions, this is \emph{not equivalent} to averaging over both $\vhat{Q}$ and $\vhat{B}$ \emph{separately}, as was stated in Ref.~[\onlinecite{Park2016}].

With these details in mind, we follow essentially the same fitting procedures as in Ref.~[\onlinecite{Rau_2016}], but using data (INS and specific heat) at zero-field and at a fixed finite field of $B = 5 \T$~\footnote{This field value was chosen as it was large enough to include significant field effects, but was not too close to either field ranges where level crossings occur.}. Rather than consider a large number of initial points in parameter space, we instead refine the values found in Ref.~[\onlinecite{Rau_2016}]. We find the refined exchanges are essentially unchanged, with
\begin{align}
\label{eq:best-global}
  J &= +0.592 \meV, &
  K &= -0.011 \meV, \nonumber \\
  \Gamma &= -0.010 \meV, &
  D &= -0.164 \meV.
\end{align}
The $g$-factors however are somewhat different, with $(g_z,g_{\pm}) = (2.72,2.30)$ compared to the values of (3.07, 2.36) we reported in Ref. [\onlinecite{Rau_2016}]. 

This model provides a good description of the excitation spectrum at both low and high fields, including the level energies and intensities, as seen in Fig.~\ref{fig:results}. One finds that two levels crossings occur, one at moderate fields $\sim 3-4 \T$ and one at larger fields $\sim 8-9\T $, consistent with $B_c^{(1)}$ and $B_c^{(2)}$ found experimentally. Both are somewhat broad due to the precise crossing being dependent on the field direction which is being averaged over. We note that whether these are true level crossings, rather than an avoided crossing, also depends on the field direction.
The agreement of the model with the experimental specific heat is not as good as with the inelastic neutron spectrum. While the agreement is reasonable above $1\K$ for $B \lesssim 5\T$ (below and a bit above the first level crossing), there are some significant disagreements at larger fields, for example at $B = 8\T$ and $B = 10\T$, as shown in Fig.~\ref{fig:results}. Like the case at zero-field, there is disagreement between the single-tetrahedron model at all fields at temperatures below $\lesssim 1\K$. As in the zero field case, we attribute this to deficiencies of the single tetrahedron model itself, not in its parametrization~\cite{Rau_2016,haku2016}.

\section{Discussion}
The agreement between the measured inelastic neutron spectrum and the theoretical model over a wide range of fields further confirms that this  pseudo-spin $1/2$
model is a good description of the single-tetrahedron physics. The disagreement for the specific heat is, however, puzzling. One should bear in mind that the energy scale associated with the specific heat is significantly lower than what is accessible in the INS data. For example, disagreement at $\sim 2\K$ or so corresponds to energy differences of order $\sim 0.2 \meV$, comparable to the energy resolution of the experiment. The specific heat data is thus a more sensitive probe of the low-energy physics of \byzo{} than the inelastic neutron scattering data.

There are several factors that could account for these discrepancies. We have checked that contributions from higher crystal field levels of the \yb{} ion do not significantly affect the results~\footnote{To gauge the importance of these terms we considered a model of the Yb$_4$ tetrahedra that included the full $J=7/2$ manifold, not just the crystal field ground doublet. In addition to a model crystal field potential, one must include multipolar exchanges between the \yb{} ions. For simplicity, we only consider rank-1 interactions, choosing their values to reproduce the fitting exchanges. This model can still be exactly diagonalized and thus its predictions can be directly compared to the model involving only the pseudo-spins. We found that the difference between these two models was entirely negligible, suggesting that the higher crystal field levels are not important at low energies.}, nor do (na\"ive) demagnetization corrections at the larger field values. Another possibility is that the field leads to modifications of the super-exchange pathways and crystal field potential (e.g. through magnetostriction) and thus leads to (weakly) field dependent values for the exchanges ($\tilde{\mat{J}}$) and $g$-factors. To explore this scenario, we refined the best fit parameters separately at each field value; while this improved the agreement somewhat at each field, there were no clear trends in the so-determined $g$-factors or the exchange values.

One potentially exciting possibility is that these discrepancies are pointing to the importance of physics beyond the single-tetrahedron model. Such effects have been discussed in explaining the low-temperature behavior at zero-field; possibilities include inter-tetrahedron interactions~\cite{Rau_2016,haku2016} or perhaps non-magnetic disorder~\cite{haku2016} splitting the doublet ground states of the tetrahedra. Given the presence of level crossings, and thus change in the ground state of the individual tetrahedra, it is possible that the relevance of the inter-tetrahedron interactions could be strongly field-dependent. For example, at small field, the magnetic moment of the (approximate) doublet ground states of the single tetrahedron are quite small, due to the \ac{DM} interaction being subdominant. The projection of the inter-tetrahedron exchange interactions~\cite{Rau_2016}
into this subspace are thus suppressed by factors of $O(\lambda^2)$. At higher fields, the ground state carries a larger magnetic moment and thus these interactions may no longer be suppressed. This is consistent with the clear manifestation of these discrepancies as one passes through the level crossings.
It is not clear if non-magnetic disorder could explain these disagreements. At low fields (below $B^{(1)}_c$), due to the small splitting of the ground doublet, one might expect non-magnetic disorder to remain effective. Once the field is large, one should expect the details of the spectrum to be less sensitive to disorder, given the single tetrahedron ground state is non-degenerate, with no nearby states that could be mixed by weak structural disorder.

In summary, the combined experimental and theoretical study reported here confirms that, at low applied fields and above 1 K, the model of independent Yb$_4$ tetrahedra provides a good description of the physics over a wide range of temperature and magnetic field.  However, below 1 K and at high fields we find significant deviations from the model.  In particular, at low temperatures, the specific heat shows a peak at $\sim$100 mK that may indicate the inter-tetrahedron interaction energy scale or may be a consequence of yet undetected disorder in \byzo{}.  On the other hand, at higher field, deviations from the model are found in both the INS data and the heat capacity data.  This discrepancy may be more consequence of inter-tetrahedron interactions between excited states becoming important, as disorder effects would not na\"ively show such a strong field dependence. 

\subsection{Acknowledgments}

This work was supported by the U.S. DOE, Office of Science, Basic Energy Sciences, Materials Sciences and Engineering Division. This research used resources at the Spallation Neutron Source, a Department of Energy (DOE) Office of Science User Facility operated by Oak Ridge National Laboratory (ORNL).  L.S.W. was supported by the Laboratory Directed Research and Development Program of ORNL, managed by UT-Battelle, LLC, for the U.S. DOE. The work at U. of Waterloo was supported by the NSERC of Canada, the Canada Research Chair program (M.J.P.G., Tier 1), the Canadian Foundation for Advanced Research and the Perimeter Institute (PI) for Theoretical Physics. Research at PI is supported by the Government of Canada through Industry Canada and by the Province of Ontario through the Ministry of Economic Development \& Innovation.   We acknowledge Tom Hogan at Quantum Design for technical assistance.

\textbf{Disclaimer}: Identification of commercial equipment does not imply recommendation or endorsement by NIST.

%

\end{document}